\documentstyle[psfig]{mn}
\newif\ifAMStwofonts
\ifoldfss

  \ifCUPmtlplainloaded \else
    \NewTextAlphabet{textbfit} {cmbxti10} {}
    \NewTextAlphabet{textbfss} {cmssbx10} {}
    \NewMathAlphabet{mathbfit} {cmbxti10} {} 
    \NewMathAlphabet{mathbfss} {cmssbx10} {} 
  \fi
  \ifAMStwofonts
    \ifCUPmtlplainloaded \else
      \NewSymbolFont{upmath} {eurm10}
      \NewSymbolFont{AMSa} {msam10}
      \NewMathSymbol{\upi}     {0}{upmath}{19}
      \NewMathSymbol{\umu}     {0}{upmath}{16}
      \NewMathSymbol{\upartial}{0}{upmath}{40}
      \NewMathSymbol{\leqslant}{3}{AMSa}{36}
      \NewMathSymbol{\geqslant}{3}{AMSa}{3E}

    \fi
  \fi
\fi 

\ifnfssone
  \newmathalphabet{\mathit}
  \addtoversion{normal}{\mathit}{cmr}{m}{it}
  \addtoversion{bold}{\mathit}{cmr}{bx}{it}
  \newmathalphabet{\mathbfit} 
  \addtoversion{normal}{\mathbfit}{cmr}{bx}{it}
  \addtoversion{bold}{\mathbfit}{cmr}{bx}{it}
  \newmathalphabet{\mathbfss} 
  \addtoversion{normal}{\mathbfss}{cmss}{bx}{n}
  \addtoversion{bold}{\mathbfss}{cmss}{bx}{n}
  \ifAMStwofonts
    \ifCUPmtlplainloaded \else
      \UseAMStwoboldmath
      \makeatletter
      \new@mathgroup\upmath@group
      \define@mathgroup\mv@normal\upmath@group{eur}{m}{n}
      \define@mathgroup\mv@bold\upmath@group{eur}{b}{n}
      \edef\UPM{\hexnumber\upmath@group}
      \new@mathgroup\amsa@group
      \define@mathgroup\mv@normal\amsa@group{msa}{m}{n}
      \define@mathgroup\mv@bold\amsa@group{msa}{m}{n}
      \edef\AMSa{\hexnumber\amsa@group}
      \makeatother
      \mathchardef\upi="0\UPM19
      \mathchardef\umu="0\UPM16
      \mathchardef\upartial="0\UPM40
      \mathchardef\leqslant="3\AMSa36
      \mathchardef\geqslant="3\AMSa3E
    \fi
  \fi
\fi 

\ifnfsstwo
  \DeclareMathAlphabet{\mathbfit}{OT1}{cmr}{bx}{it}
  \SetMathAlphabet\mathbfit{bold}{OT1}{cmr}{bx}{it}
  \DeclareMathAlphabet{\mathbfss}{OT1}{cmss}{bx}{n}
  \SetMathAlphabet\mathbfss{bold}{OT1}{cmss}{bx}{n}
  \ifAMStwofonts
    \ifCUPmtlplainloaded \else
      \DeclareSymbolFont{UPM}{U}{eur}{m}{n}
      \SetSymbolFont{UPM}{bold}{U}{eur}{b}{n}
      \DeclareSymbolFont{AMSa}{U}{msa}{m}{n}
      \DeclareMathSymbol{\upi}{0}{UPM}{"19}
      \DeclareMathSymbol{\umu}{0}{UPM}{"16}
      \DeclareMathSymbol{\upartial}{0}{UPM}{"40}
      \DeclareMathSymbol{\leqslant}{3}{AMSa}{"36}
      \DeclareMathSymbol{\geqslant}{3}{AMSa}{"3E}
    \fi
  \fi
\fi 

\ifCUPmtlplainloaded \else
  \ifAMStwofonts \else 
    \def\upi{\pi}
    \def\umu{\mu}
    \def\upartial{\partial}
  \fi
\fi

\def\et{{\it et al.\ }}
\title[Very High State Geometry]
{The very high state accretion disc structure from the galactic black
  hole transient XTE~J1550-564}
\author[Aya Kubota, Chris Done]
{Aya Kubota$^{1}$, 
Chris Done$^{2}$ \\
$^1$Cosmic Radiation Laboratory,
                 Institute of Physical and Chemical Research,\\
2-1 Hirosawa, Wako-shi, Saitama, 351-0198, Japan; aya@crab.riken.jp\\
$^2$Department of Physics, University of Durham, South Road,
Durham, DH1 3LE, England; chris.done@durham.ac.uk}

\date{Accepted 2004 * **.
      Received 2003 * **;
      in original form 2003 * **}
\begin{document}

\maketitle

\begin{abstract}

The 1998 outburst of the bright Galactic black hole binary system
XTE~J1550-564 was used to constrain the accretion disc structure in
the strongly Comptonised very high state spectra. These data show that
the disc emission is not easily compatible with the constant area
$L\propto T^4$ behaviour seen during the thermal dominated high/soft
state and weakly comptonised very high state.
Even after correcting for the effects of the scattering
geometry, the disc temperature is always much lower than expected for
its derived luminosity in the very high state. The simplest
interpretation is that this indicates that the optically thick disc is
truncated in the strongly Comptonised very high state, so trivially
giving the observed continuity of properties between the low/hard and
very high states of Galactic black holes.

\end{abstract}



\section{Introduction}

Accreting black holes in our Galaxy show a large variety of X-ray properties, 
which can be classified into distinct spectral states (Tanaka \& Lewin 1995; 
McClintock \& Remillard 2003, hereafter MR03).  In the high/soft state, 
typically seen at luminosities above $\sim 5-10$ per cent of Eddington, the 
energy spectrum is dominated by a soft thermal component of temperature 0.4--
1.5~keV. This is well described by the standard accretion model of emission from 
an optically thick and geometrically thin disc (Shakura \& Sunyaev 1973, 
hereafter SS73). Where the source varies, the temperature and luminosity of this 
disc component change together in such a way as to indicate that the size of the 
emitting structure remains approximately constant i.e. $L\propto T^4$ (Ebisawa 
\et ~1991; 1994; Kubota, Makishima \& Ebisawa 2001; Kubota \& Makishima 2004; 
hereafter KM04; Gierlinski \& Done 2004; hereafter GD04).  Excitingly, this 
gives an observable diagnostic of the mass and spin of the black hole, assuming 
that this constant size scale is set by the innermost stable orbit around the 
black hole.

These results are actually very surprising in the context of
the SS73 disc equations. The predicted disc structure is violently
unstable at such luminosities, where radiation pressure dominates, yet
these disc dominated spectra show very little short time-scale
variability. Also, the disc spectrum should {\em not} be a simple sum
of blackbody spectra at these high temperatures, as electron
scattering dominates over true absorption. In the X-ray band, this
difference in spectral shape can be roughly described as a standard
disc spectrum, but of a higher temperature than the effective
(blackbody) temperature (Shimura \& Takahara 1995; Merloni et
al. 2000). However, this colour temperature correction factor can
change with luminosity, so is {\em not} predicted to produce
a simple $L\propto T^4$ relation (GD04).
	
Real discs are then observed to be much simpler than the SS73 predictions, 
pointing to the inadequacies of the {\em ad hoc} $\alpha$ viscosity prescription 
and motivating further studies of the disc structure which results from the 
physical (magnetic dynamo) viscosity mechanism in radiative discs (e.g. Turner 2004).  This observed simplicity is reserved only for the high/soft state data, 
where the disc spectrum is dominant. Accreting black holes can also show a 
rather different type of spectrum at high luminosities, the so-called very high 
state termed the steep power-law state by MR03), characterized by a very strong 
(roughly) power-law component (Miyamoto \et 1991). 
This is normally steep, with photon spectral 
index $\Gamma > 2.2$, rather different to the standard low/hard state spectra 
seen usually at much lower luminosities (e.g. the reviews by Tanaka \& Lewin 
1995; MR03). In this very high state,  the constant area inferred from the 
relation between observed temperature and disc luminosity breaks down (Ebisawa 
\et  1994; Kubota \et  2001; KM04; MR03).

To some extent this breakdown is expected. The strong high energy emission shows 
that a large fraction of the accretion energy is dissipated in some non-disc 
structure (corona/jet?), so the disc structure should be different to that seen 
when the emission is dominated by the thermal disc spectrum (e.g. Svensson \& 
Zdziarski 1994). Also, the intense X-rays may illuminate the disc, again 
changing its structure (e.g.\ Nayakshin, Kazanas \& Kallman 2000). However, 
Kubota et al (2001) showed that for very high state spectra where the high 
energy emission was less than $\sim 50$ per cent of the total flux (which they 
termed the anomalous spectra), then the constancy of disc area could be 
(approximately) recovered by careful modelling of the spectrum, and by 
correcting the observed disc luminosity to include the Compton scattered photons 
(KM04).

Here we investigate the extremely Comptonised very high state spectra
from the 1998 outburst of XTE J1550-564, where the steep power
law tail dominates the total emission. This source has a superluminal
jet (Hannikainen \et 2001) and is a confirmed black hole with mass of
8.4--11.2 $M_\odot$, at a distance of 5~kpc and binary inclination
angle of $70^\circ$ (Orosz \et ~2002).

We show that these extremely Comptonised very high spectra are qualitatively and 
quantitatively different to the spectra with a weaker Comptonised tail. They are 
not easily compatible with a simple Comptonising corona above an untruncated 
disc, though such a geometry may be possible with a complex corona, with strong 
radial gradients in the optical depth. However, the simplest solution is that the 
disc in these Compton dominated spectra is truncated at larger radii than the 
minimum stable orbit around the black hole.

\section{Spectral evolution of XTE~J$1550-564$ in the 1998 outburst}

\begin{figure}
\centerline{\psfig{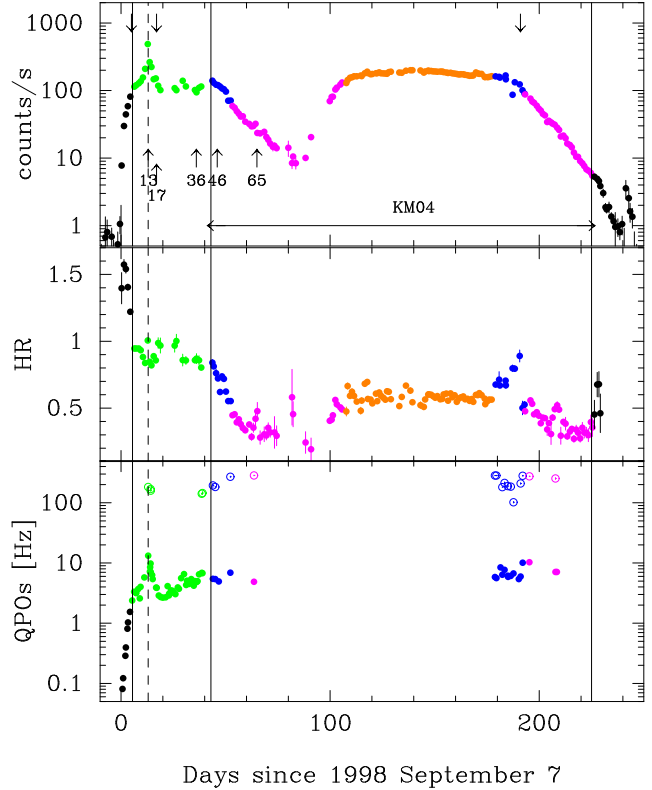}}
\caption{The {\it RXTE} ASM light curves of XTE~J$1550-564$. The top panel
shows the 1.5--12~keV ASM count rate, while the middle panel shows the
ASM hardness ratio (5--12~keV/3--5~keV).
The bottom panel shows the high (open circles) 
and low (filled circles) frequency QPOs, taken from Remillard \et ~(2002).
Harmonics of the low frequency QPOs, reported by Remillard \et~(2002), 
are not shown here. 
The times of the simultaneous ASCA/{\it RXTE} observations are indicated with 
down-arrows in the top panel, while up-arrows indicate the times of 
the spectra shown in Fig.~\ref{fig:uspec}. 
Vertical solid lines show days 5, 43, and 225 since the beginning of 
the outburst, and the vertical dashed line indicates the 
position of the outburst peak. 
Data points with blue, magenta, and orange correspond to the anomalous regime 
(or weak very high state in text), standard regime, and apparently standard regime defined by KM04, respectively. Green data points correspond to
the strong very high state which is introduced in this paper.}
\label{fig:asm}
\end{figure}

The transient black hole binary XTE~J$1550-564$ was discovered on 1998 September 
7 by {\it RXTE} ASM (Smith 1998) and {\it CGRO} BATSE (Wilson \et  1998).  This 
source showed four subsequent outbursts since its discovery, but the first 
outburst was the brightest and best covered by  {\it RXTE} pointing 
observations. The data were analyzed by many authors including Sobczak \et 
~(1999; 2000ab), Wilson \& Done (2001), Homan \et  (2001), and KM04.


Figure~\ref{fig:asm} shows the 1.5--12~keV light curve and hardness ratios of 
the first outburst, obtained with {\it RXTE} ASM, together with the frequency of 
the main QPO (where seen) from Remillard \et (2002). Clearly the source hardness 
changes dramatically during the outburst, as does the QPO frequency. These are 
generally correlated, so that spectral states can be defined either from 
spectral or power spectral characteristics (van der Klis 2000; MR03). 

Here we briefly summerise the main features of the source behaviour. 
The marked softening of the spectrum in the first 
5-10 days after the source
was detected is associated with a state transition from 
the low/hard state (first five days) into the
very high state (Wilson \& Done 2000; Cui \et 1999; Sobczak \et 1999).
For about a month (between days 5--52: Fig.~1),
the source showed strong QPOs and 
a strong hard spectral component (e.g., Sobczak \et  2000a), so 
can be classified as in the very high state during the whole of this time.
The QPO frequency changes
quite dramatically during the first $\sim$ 40
days of the outburst, but then stabilizes from day 40-52. The type of QPO
also changes at day 40 (from C/C' to B: see Remillard \et ~2002).

After the day 52, the luminosity and hardness both decreased 
as the source made a 
transition to the classic high/soft disc dominated (termed thermal
dominated: MR03) state. 
Generally the QPO is not seen, but when it is detected, 
it is of different type (A, very weak: Remillard \et 2002) to that seen 
previously. Most of the time until the end of this first outburst, the source 
stayed in this disc dominated high/soft state 
as the luminosity changed, giving progressively higher/lower disc 
temperatures and so higher/lower hardness ratios. There is just one section 
(days 180-200) where the hard power law flux again increased relative to the 
disc emission to a level comparable to that seen towards the end of the first 
very high state (days 40-52) and where the QPOs (mainly type B) reappear.
Hence Homan \et (2001)
identify these data as very high state.

The luminosity and temperature of the disc component from day 40 onwards is 
always consistent with optically thick material extending down to the last 
stable orbit around the black hole (KM04, see also Figs 4b-d). 
In this paper, we concentrate our discussion on the very high state spectra 
where the coronal luminosity is $>$ 50\% of the total i.e. days 5--40 of the 
outburst. We will show that these data are most easily described if the inner 
disc does NOT extend down to the last stable orbit, as also indicated by the QPO 
frequency. Hereafter, we distinguish 
these data from the other very high states in which the power law component is 
weaker (days 40--52, 180--200), and call it the "strong very high state".  In 
Table~\ref{tab:name}, we summarize the correspondence of the spectral states we 
used in this paper to the state defined in other literatures.  

\begin{table*}
\centering
\begin{minipage}{175mm}
\caption{Correspondence of the definition of the state name to those in other papers.}
\label{tab:name}
\begin{tabular}{ccccccc}
\hline \hline
This paper  & KM04 & classic classification & MR03 & QPO type $^\dagger$
 & corresponding obs.  & Remarks $^\ddagger$\\
\hline
low/hard & --- &low/hard (low) & hard &C or C' &  0--5 & thermal IC\\
strong very high & --- & very high (high) & SPL & C or C' & 5--40& thermal IC\\
weak very high & anomalous  &  very high (high) & SPL & B$^\ast$ & 40--52, 180--200& thermal IC\\
standard high/soft&  standard  &high/soft (high) & TD & A or none &52--100, 200--220& dominant disc + PL\\
(apparently standard)$^\S$& apparently standard  & high/soft (high) & TD & none &100--180 & dominant disc + weak PL\\
\hline
\multicolumn{7}{l}{$^\dagger$ By referring to Remillard \et 2002}\\
\multicolumn{7}{l}{$^\ddagger$ Characteristics of X-ray spectrum}\\
\multicolumn{7}{l}{$^\ast$ Majority of QPOs are found in type B, while 
type A and B QPOs are reported on day 192 and day 187, respectively.}\\
\multicolumn{7}{l}{$^\S$ Not described in the text.}\\
\end{tabular}
\end{minipage}
\end{table*}

\begin{figure}
\centerline{\psfig{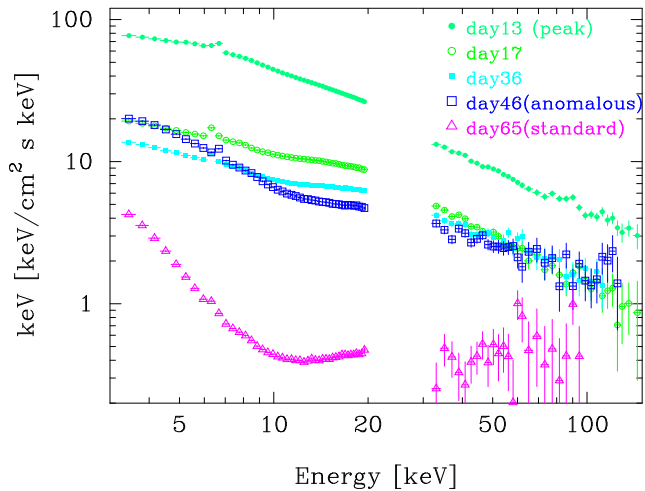}}
\caption{Typical PCA and HEXTE spectra of XTE~J$1550-564$
taken from times as indicated in Fig. 1.
The data is unfolded with the best fit model fit to the PCA+HEXTE data
(see \S 4.2).}
\label{fig:uspec}
\end{figure}

\section{Data reduction}

We used all the data from the first outburst
except for the first 5 days where the source is in the low/hard state.
Good PCA and HEXTE data were selected and processed, 
following the same procedures as in KM04 and WD01, respectively.

For the PCA data reduction we used top and mid layer from all available
units, using standard exclusion criteria (target elevation
less than $10^\circ$ above the Earth's limb,
pointing direction was more than $1^{\prime}\!.2 $ from the target,
data acquired within 30 minutes
after the spacecraft passage through South Atlantic Anomaly).
The standard dead time correction procedure was applied to the data.
The PCA background was estimated for each observation,
using the software package {\it pcabackest} (version 2.1e),
supplied by the {\it RXTE} Guest Observer's Facility at NASA/GSFC. 
The PCA response matrix was made for each observation by utilizing
the latest version (8.0) of the software package {\it pcarsp}. 
In order to take into account the calibration uncertainties,
0.5\% systematic errors are added to each bin of the PCA spectra.
A more conservative 1\% systematic error does not qualitatively change the
results, but gives very small values of $\chi^2/{\rm dof}$.
The 3--20~keV data are used in this paper, since there is sometimes residual 
structure in the 20--35 keV range associated with the Xe-K edge at $\sim$30~keV.

The HEXTE consists of two independent clusters (Cluster 0 and 1) of four 
NaI(Tl)/CsI(Na) phoswich scintillation counters (Rothschild \et  1998). The 
HEXTE data were 
selected in the same way as that of the PCA. Since one detector 
on Cluster 1 of the HEXTE has lost its spectral capability, we use 
only {\it Standard 
mode} data from Cluster 0. The HEXTE background 
was subtracted by sequential rocking of the two clusters on and off of the 
source position. We used the software package called {\it hxtback} for splitting 
the data and the background. 
The HEXTE data  below $30~$~keV were not used in the spectral fit 
because of response uncertainties associated with the Xe-K edge\footnotemark
\footnotetext{
see http://mamacass.ucsd.edu/hexte/calib/README.hexte\_97mar20}.
Systematic errors of 0.5\% are also added to 
each spectral bin of the HEXTE,
though these are much smaller than statistical errors of the HEXTE data in the range of 
30--200~keV so make no difference to the spectral fits. 
Figure~\ref{fig:uspec} shows the range of spectral shapes seen by 
the PCA and HEXTE data over the course of the outburst, taken from days of 13, 
17, 36, 46, and 65 as denoted by the up-arrows in Fig.~\ref{fig:asm}.

ASCA observations of this source were performed three times, on 1998
September 12, 24, and 1999 March 17, as indicated with down arrows in
Fig.~\ref{fig:asm}.  There are simultaneous {\it RXTE} observations
corresponding to each ASCA observation.  The first simultaneous
observation was just on the boundary of the low and very high state,
while the third one was in the weak very high state (days 180-200).  In
this paper we are most interested in the strong very high state, so we choose
to use only the second ASCA observation.
The ASCA GIS events were extracted from a circular region 
of $6'$ radius centered on the image peak, 
after selecting good time intervals in a standard procedure.
A net exposure of 2.4~ks was obtained for this simultaneous pointing.
Dead time fractions are determined from count rate monitor data
(Makishima \et  1996) as 85.6\% for GIS2 and 87.3\% for GIS3.
Systematic errors of 1\% are added to each spectral bin. 
We only use the GIS data 
since the SIS is strongly affected by pileup.

The well known difference in calibration between the GIS and the PCA has been 
fixed in the PCA response matrices in HEASOFT 5.2.
Thus the data are fitted simultaneously in this paper. 

\begin{table*}
\centering
\begin{minipage}{175mm}
\caption{Best fit parameters of XTE J$1550-564$ }
\label{tab:rxte}
\begin{tabular}{cccccccccc}
\hline \hline
day & $T_{\rm in}$ & $\Gamma$ & $T_{\rm e}$ & 
$L_{\rm disk}$\footnote{The disc bolometric luminosity in the unit of 
$10^{38}~{\rm erg~s^{-1}}$.}
& $L_{\rm thc}$\footnote{The isotropic thcomp luminosity in the range of 
0.01--100~keV in the unit of $10^{38}~{\rm erg~s^{-1}}$.} 
& $L_{\rm pow}$\footnote{The isotropic power-law luminosity in the range of 
1--100~keV in the unit of $10^{38}~{\rm erg~s^{-1}}$.} 
&   smedge &line\footnote{$\sigma$ is fixed at 0.1~keV.}
 & $\chi^2/{\rm dof}$\\
 & keV & $\Gamma_{\rm thc}$ & keV 
 &$(N_{\rm disk})$\footnote{0.01--100~keV photon flux in the unit of 
 photons~${\rm s^{-1}cm^{-2}}$.}  
 &$(N_{\rm thc})^e$
 && & &\\
\hline \hline
\multicolumn{10}{c}{MCD + power-law (PCA)}\\
\hline \hline
17 & $0.38^{+0.01}_{-0.02 }$&$2.46\pm0.01$ & ---  & 1.98 &---  &3.24  &   $E=8.0^{+0.3}_{-0.2}$~keV  &$E=6.44\pm0.08 $ keV & 24.0/36\\
& & && & & &    $\tau_{\rm max}=0.17^{+0.05}_{-0.03}$ & EW=$67\pm12$~eV&  \\
& & && & & &   width$=1.3^{+0.6}_{-0.5}$~keV  &  &\\
\hline
36 & $0.90^{+0.08}_{-0.12}$&$2.33\pm0.03$ & ---  & 0.56 &---  &1.97  &  $E= 8.1\pm0.4$~keV  &$E=6.48^{+0.12}_{-0.05}$ keV & 24.0/36\\
& & && & & &    $\tau_{\rm max}=0.5^{+0.5}_{-0.3}$ & EW=$64^{+15}_{-19}$~eV&  \\
& & && & & &   width$=4^{+6}_{-2}$~keV  &  &\\
\hline
46 & $1.05\pm0.02$&$2.42\pm0.03$ & ---  & 1.70 &---  &1.76  &   $E=8.6^{+0.3}_{-0.2} $~keV  &$E=6.7^{+0.2}_{-0.1}$ keV & 17.0/36\\
& & && & & &    $\tau_{\rm max}>2.0_{-1.3}$ & EW=$45\pm12$ eV&  \\
& & && & & &   width$=18^{+3}_{-12}$~keV  &  &\\
\hline
65 & $0.72\pm0.01$&$2.07\pm0.04$ & ---  & 1.05 &---  &0.12  &   $E=8.6\pm0.2 $~keV  &$E=6.5\pm0.2$ keV & 37.2/36\\
& & && & & &    $\tau_{\rm max}>2.0_{-0.5}$ & EW=$68\pm16$ eV&  \\
& & && & & &   width$=8^{+1}_{-2}$~keV  &  &\\
\hline \hline
\multicolumn{10}{c}{MCD + power-law + {\rm thcomp} (PCA \& HEXTE)}\\
\hline \hline
17 & $0.65^{+0.08}_{-0.04}$&$2.32\pm0.03$ &  $13\pm2$  & 1.21 &2.50&0.69 &   $E=7.9^{+0.4}_{-0.3}$ keV  &$E=6.40\pm0.08$ keV &35.9/72\\
& & && (29.9) &(42.9) & & $\tau_{\rm max}=0.35\pm0.13$ & EW=$77^{+17}_{-9}$ eV&  \\
& & && & & &   width$=3\pm2$~keV  &  &\\
\hline
36 & $0.78^{+0.05}_{-0.06}$&$2.25\pm0.04$ &  $22^{+30}_{-7}$  & 1.01 &1.51 &0.51 &   $E=7.9^{+0.4}_{-0.3}$ keV  &$E=6.4\pm0.1$ keV &49.4/72\\
& & &&(22.2) &(20.2) & &    $\tau_{\rm max}=1.2^{>+0.8}_{-0.7}$ & EW=$65^{+18}_{-13}$ eV&  \\
& & && & & &   width$=10^{+8}_{-6}$~keV  &  &\\
\hline
46 & $0.98\pm0.02$&$2.40\pm0.08$ &  $19^{>+181}_{-8}$  & 2.13 &1.08 &0.72 &   $E=8.6^{+0.2}_{-0.1}$ keV  &$E=6.7^{+0.2}_{-0.1}$ keV &50.8/72\\
& & &&(38.2) &(13.8) & &    $\tau_{\rm max}>2.0_{-1.0}$ & EW=$54\pm12$ eV&  \\
& & && & & &   width$=15^{+2}_{-7}$~keV  &  &\\
\hline
\multicolumn{10}{l}{Errors represent 90\% confidence level for one free parameter i.e.
$\Delta\chi^2=2.7$.}\\
\multicolumn{10}{l}{Normalization factors with the PCA data are used to calculate luminosities. }\\
 \end{tabular}
\end{minipage}
\end{table*}

\section{Analyses of the multiple {\it RXTE} data}

\subsection{Simple fits to the PCA data}

In order to briefly characterize the spectral behavior 
throughout the outburst, we performed
spectral fitting to the 3--20~keV PCA data with the canonical
multicolour disc model (hereafter MCD model, 
{\sc diskbb} in XSPEC; Mitsuda \et  1984)  plus
a power-law model. 
The power-law component is modified by an absorption edge around 7--9~keV described
by a smeared edge model ({\sc smedge} in XSPEC; Ebisawa \et ~1994),
and a narrow Gaussian 
is also added for the Fe-K line by constraining its central energy in the 
range 6.2--6.9~keV. 
The data of days 80--165 cannot constrain the power-law photon index
so the index was fixed at 2.0, representing the mean HEXTE slope (Sobczak \et
2000b; KM04).
We fix the hydrogen column at 
$N_{\rm H}=7\times10^{21}~{\rm cm^{-2}}$ as indicated by the ASCA data
(see table 3), but otherwise the analysis is the same as in KM04.

\begin{figure}
\centerline{\psfig{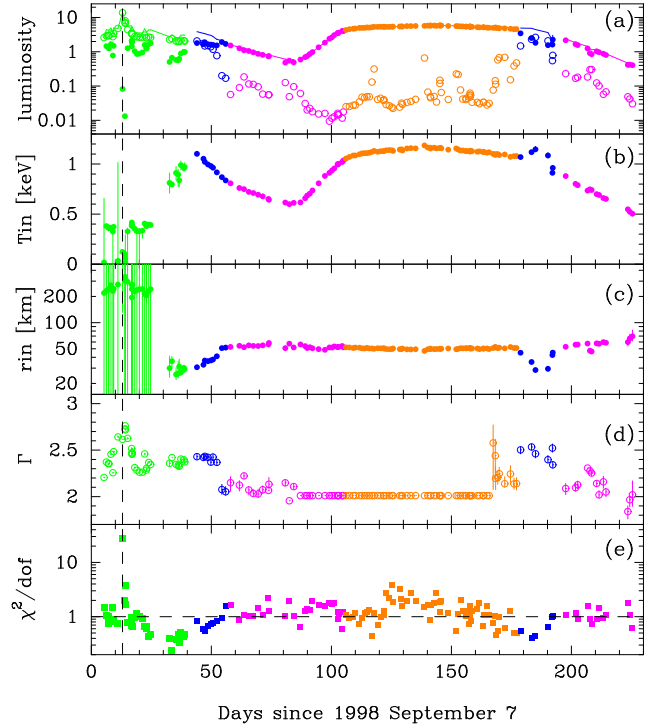}}
\caption{Evolution of the spectral parameters of XTE~J$1550-564$ based 
on the PCA data fitted by a simple MCD plus power-law model (see \S~4.1). 
(a)Time histories of $L_{\rm disk}$ (filled circles), 
$L_{\rm pow}$ (solid circles), and $L_{\rm tot}$ (solid line), 
all in units of 
$10^{38}\cdot {D_5}^2~{\rm erg~s^{-1}}$ and assuming
$i=70^\circ$ for the disc emission. (b)--(e) Those of $T_{\rm in}$, 
$r_{\rm in}$, $\Gamma$, and $\chi ^2/{\rm dof}$, respectively. 
}
\label{fig:evolution1}
\end{figure}

Except for the data at the peak (day~13),
this model fits the data adequately, and Table~\ref{tab:rxte} gives the best fit
parameters for the spectra shown in Fig.~\ref{fig:uspec}.
Figure~\ref{fig:evolution1} shows the evolution of the spectral parameters
as a function of time (as in Fig.~2 of KM04 but including the strong
very high state data of the first 5--40 days as well). Panel (a) shows
the bolometric luminosity of the disc, $L_{\rm disk}$, 1--100~keV
power law luminosity, $L_{\rm pow}$ and total $L_{\rm tot}\equiv L_{\rm
disk}+L_{\rm pow}$. Here, $L_{\rm disk}$ is calculated by referring to 
two parameters of the MCD model, inner disc temperature $T_{\rm in}$
(panel b) and an apparent inner disc radius $r_{\rm  in}$ (panel c),
as $L_{\rm disk}=4\pi r_{\rm in}^2 \sigma T_{\rm in}^4$.

\begin{figure*}
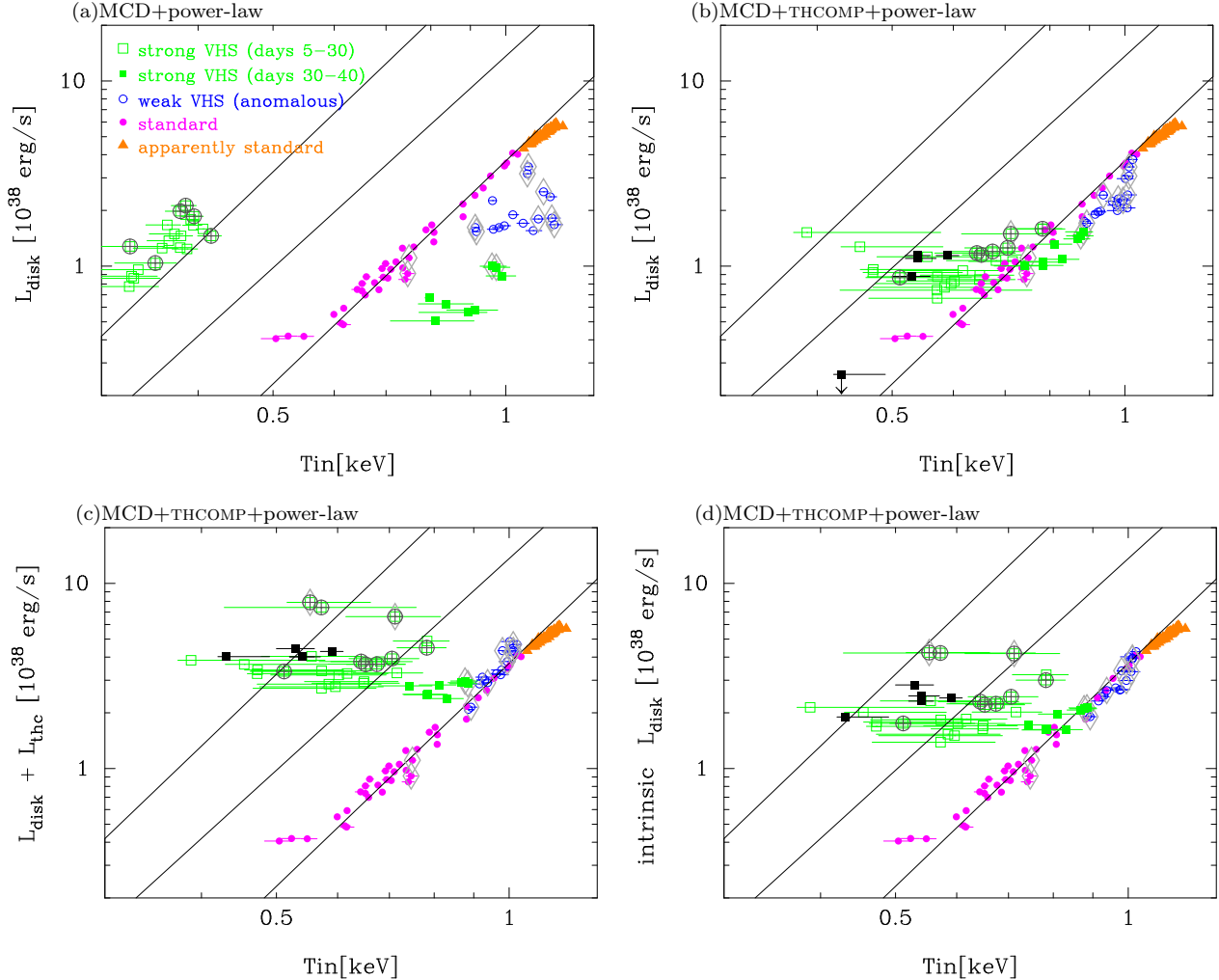

\begin{minipage}{240pt}
\hspace{30pt}(a)MCD+power-law
\centerline{\psfig{file=t-ld_mp.ps,width=230pt}}
\end{minipage}
\begin{minipage}{240pt}
\hspace{30pt}(b)MCD+{\sc thcomp}+power-law
\centerline{\psfig{file=t-ld_thc.ps,width=230pt}}
\end{minipage}
\vspace*{10pt}

\begin{minipage}{240pt}
\hspace{30pt}(c)MCD+{\sc thcomp}+power-law
\centerline{\psfig{file=t-ld+lthc_thc.ps,width=230pt}}
\end{minipage}
\begin{minipage}{240pt}
\hspace{30pt}(d)MCD+{\sc thcomp}+power-law
\centerline{\psfig{file=t-ld_cor.ps,width=230pt}}
\end{minipage}
\caption{The disc luminosity plotted against the observed inner disc
  temperature, $T_{\rm in}$. (a) $L_{\rm disk}$ is 
taken from the  MCD plus power-law model fits to the PCA data (\S~4.1). 
(b) $L_{\rm disk}$ is 
taken from the three-component model fits to the PCA+HEXTE data, including 
the {\sc thcomp} model to describe  the thermal Comptonisation (\S~4.2). 
(c) $L_{\rm disk}+L_{\rm thc}$  from the three-component model  (\S~4.2).
(d)) $L_{\rm disk}^{\rm int}$ estimated by conservation of photon
  number (\S~5). In all panels the solid lines show lines of constant apparent inner
  disc radius of 59, 100 and 200~km, with colours indicating different spectral states
  (see Fig 1. and table 1). Additionally, points highlighted by a large circle with a 
  cross are those where there is also strong radio emission from the jet, while 
  diamonds show the ones where the high frequency QPO is detected. Panels (b)-(d)
  also include the data from the five different fits to the ASCA-PCA-HEXTE spectrum
  (Table 3) as black squares. 
 In panel (a), some data points around the peak are not included, and 
in panels (b)--(d), the data point at the peak is not included. }
\label{fig:t-l}
\end{figure*}

Figure~\ref{fig:t-l}a plots the derived disc luminosity against
inner disc temperature for these fits.  Clearly, $L_{\rm disk}\propto
T_{\rm in}^4$ in the standard high/soft state (magenta filled circles),
indicating a constant area emitting region. Only 
the very lowest luminosity points deviate from this line, but the errors
on these are fairly large as the low temperature disc is rather poorly
covered by the PCA bandpass. 
When we fit these data points with a power-law, 
we obtain an index of $3.9^{+0.1}_{-0.2}$ with 
a normalization at $T_{\rm in}=1$~keV 
of $L_{\rm disk}=3.55\pm0.08~{\rm erg~s^{-1}}$. 
It is very tempting to
identify the source of this markedly constant behaviour with the 
constant area
defined by a disc extending down to the innermost stable orbit. When
the apparent inner disc radius of $r_{\rm in}=53$~km derived from the fits is
corrected for the stress--free inner boundary condition (Kubota et
al. 1998) and colour temperature correction of 1.7 then this gives 
an estimate for the true inner disc radius of $R_{\rm in}=63$~km
with a typical error of 10~\%.
This value is slightly smaller than 75--100~km, which is the innermost 
stable orbit for a non-spinning black hole of 8.4--11.2~$M_\odot$.
Therefore, XTE~J$1550-564$ is consistent with a moderately rotating black hole, 
while uncertainties of the source distance and determination 
of $\kappa$ can cancel this difference.

%
%

The behaviour of the disc component in the rest of the data
is rather different.  Fig.~\ref{fig:t-l}a shows that the strong very
high state spectra (green open squares, days 5--30) have a much lower
temperature disc than expected for the disc luminosity. Conversely the
disc in the weak very high states 
has the disc at somewhat higher temperature than expected from its luminosity. 
In other words, the disc inner radius of the strong very high state 
appears at much larger values than that disc dominated regime, while 
that of the weak very high state is apparently found at smaller values.
Inspection of Fig.~\ref{fig:evolution1}(b)--(c) shows that this
change in behaviour happens around day 30.

\subsection{Fitting with the picture of the strong disc Comptonisation}

The key characteristic of all the non high/soft state data is
that the Comptonised tail is much more important in the spectrum (see
Fig.~\ref{fig:uspec}), so details of how it is modeled become
important.  A power law approximation for the Comptonised spectrum is
inadequate, as real Compton spectra show a low energy turnover close
to the seed photon energy (KM04; Done, Zycki \& Smith 2002).
We replace the power law with 
a proper thermal Comptonisation model 
({\sc thcomp}\footnote{http://www.camk.edu.pl/\symbol{"7E}ptz/relrepr.html}
in XSPEC: Zdziarski, Johnson \& Magdziarz 1996)
which is parameterized by asymptotic power law index,
$\Gamma_{\rm thc}$, and electron temperature, $T_{\rm e}$. We assume that
the seed photons are from the observed MCD component,
so $T_{\rm e}$ is the only additional free parameter
compared to the previous power law fits.

We use the PCA and HEXTE data in order to better constrain 
$T_{\rm e}$, but this broad bandpass shows that the spectra
are rather more complex than a single thermal comptonisation component
(and its reflected emission). While the data below 20 keV can be
dominated by the disc and a cool thermal Compton component, the higher
energy data show a quasi-power law tail (Gierlinski \et  1999;
Zdziarski \et  2001; WD01; Kubota \et  2001;
Gierlinski \& Done 2003; KM04). We follow KM04 and model this
additional component as a power law with photon index $\Gamma$ fixed
at 2.0 which is the average index 
of the high energy emission in the standard high/soft state (e.g. fits to HEXTE data:
Sobczak \et 2000b), and fit all the PCA+HEXTE data with this three
component continuum model; MCD plus thermal comptonisation ({\sc thcomp}) 
plus power law, with the gaussian line and smeared edge to mimic reflection.
All the model parameters are constrained to be
the same between the PCA and the HEXTE data
except for a normalization factor.

This three component model successfully reproduced all the spectra, and the 
electron temperature of the Comptonised component was found at 10--30~keV in all 
the very high state data, except for that at the peak (day 13) where the 
temperature is considerably higher at $\sim 60$~keV. This means that the thermal 
rollover in the spectrum is not so obvious in the PCA-HEXTE bandpass. 
Nonetheless, the complex continuum model (thermal plus non-thermal) is still 
strongly preferred by the data compared to a single Comptonised component. Thus 
it seems likely that the peak spectrum is simply a more extreme version of 
the surrounding very high state spectra. Alternatively it could contain an
additional component such as an X-ray contribution from the jet, 
whose radio emission suddenly increased at this time (day~13), peaked on day~15, 
and decayed to quiescent level through day~18 (e.g., Wu \et ~2002).

\begin{figure}
\centerline{\psfig{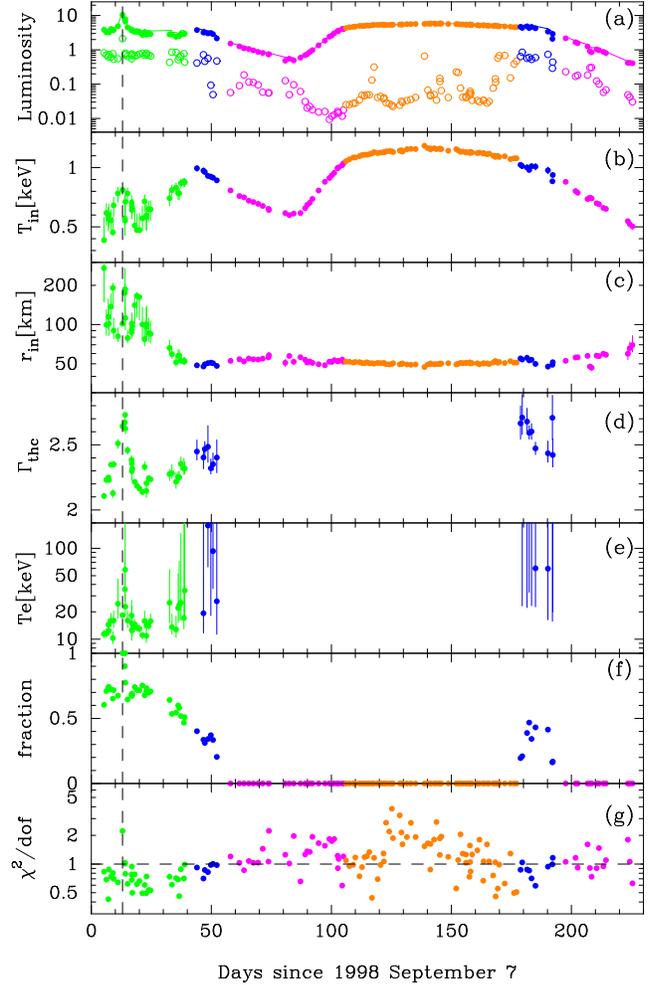}}
\caption{Same as Fig.~\ref{fig:evolution1}, but based on the three
  component model fits to the PCA+HEXTE data (\S 4.2).
The data points after day 52 are the same as in Fig.~\ref{fig:evolution1}.
(a)Time history of $L_{\rm disk}+L_{\rm thc}$ is plotted with filled circle
instead of $L_{\rm disk}$ for the very high state data, where 
$L_{\rm thc}$ is estimated under an assumption of isotropic emission.
(b)--(f) Those of $T_{\rm in}$, 
$r_{\rm in}$, $T_{\rm e}$, $\Gamma_{\rm thc}$, 
$L_{\rm thc}/(L_{\rm disk}+L_{\rm thc})$, 
and $\chi^2/{\rm dof}$, respectively. Panels (a) and  (c) are based on the normalization 
factor obtained with the PCA data. 
}
\label{fig:evolution2}
\end{figure}

Figure~\ref{fig:evolution2} shows the evolution of the spectral
parameters with time, while the resulting luminosity/temperature
diagram is shown in Fig.~\ref{fig:t-l}b.  The change in derived disc
parameters for spectra shown in Fig.~\ref{fig:uspec} are given in Table~\ref{tab:rxte}.
The disc temperature
increases dramatically for the strong very high state spectra 
(days 5--30, green open squares), while for the weak very high state
spectra (days 40--52 shown as blue open circles), 
it is the disc luminosity which increases. The
net effect is to pull all the Comptonised spectra much closer to the
constant radius behavior seen in the standard high/soft state (magenta filled
circles: Fig.~\ref{fig:t-l}b).  This looks very promising for models
in which the disc structure remains rather stable in all the high
state spectra, with constant apparent inner disc radius.

However, the Comptonised luminosity is indeed very large 
compared to that of the disc in the very high state, 
making it likely that the corona covers a fairly large
fraction of the disc. In this case the corona is in the line of sight
and intercepts some of the disc photons, so the observed disc luminosity
underestimates the intrinsic disc luminosity, 
$L_{\rm disk}^{\rm int}$. The energy transfer by Compton scattering
can be characterised by the Compton $y$ parameter, which is 
$\sim 4\tau^2 kT_e/mc^2$ (e.g., Rybicki \& Lightman 1979). 
If this is small then 
$L_{\rm disk}^{\rm int}$ is roughly consistent with 
$L_{\rm disk}+L_{\rm thc}$ (Kubota \et  2001; KM04).
Figure~\ref{fig:t-l}c shows this luminosity plotted against $T_{\rm
in}$. As shown by KM04 this puts the weak very high state 
(anomalous regime ; days 40--52)
data onto the same constant radius line as the standard regime data,
while the strong very high state lies significantly above these
points.

For the strong very high state, the Compton energy boost is
{\em not} negligible.  Instead, we estimate $L_{\rm disk}^{\rm int}$ 
by conservation of photon number, as the photons in the Comptonised
spectrum came originally from scattering of seed photons from the disc.
Figure~\ref{fig:geo1} shows
two simple corona geometries for the very high state, a sphere and a
slab, respectively, above an untruncated disc. 
The intrinsic number of disc photons, $N_{\rm disk}^{\rm int}
= N_{\rm disk} + a N_{\rm thc}$, 
where values of $a$ are $2\cos i$ and 1, for spherical and slab
geometries, respectively (KM04), assuming that the seed photons
from the disc have the same disc blackbody spectrum as is observed. 
Thus, the intrinsic disc emission can be estimated by 
\begin{eqnarray}
L_{\rm disk}^{\rm int}&=&L_{\rm disk}\cdot N_{\rm disk}^{\rm int}/N_{\rm disk}\\
&=&b T_{\rm in} \cdot N_{\rm disk}^{\rm int}
\end{eqnarray}
Here, the value of $b$ is defined as $bT_{\rm in}N_{\rm disk}=L_{\rm disk}$, and 
via equation (A1) in KM04, 
it is given as
\begin{equation}
b=1.31\times10^{-9}\cdot \frac{2\pi D^2}{\cos i}~~~{\rm erg~keV^{-1}photons^{-1}cm^2}~~,
\end{equation}
for the bolometric (0.01--100~keV) photon fluxes, $N_{\rm disk}$ and
$N_{\rm thc}$.  

Figure~\ref{fig:t-l}d shows this estimate for $L_{\rm
disk}^{\rm int}$ (calculated for a spherical geometry) plotted against
$T_{\rm in}$, while Fig.~\ref{fig:evolution2}c shows the conversion of
$L_{\rm disk}^{\rm int}$ into apparent inner disc radius. 
It is clear that the weak very high state (anomalous regime) data are now
easily consistent with the same constant $r_{\rm in}$ seen in the standard 
high/soft data, but the strong very high state, where the
Comptonised luminosity is more than $\sim 50$\% of the total flux, are
well above the line. This is {\em highly unlikely} to be due to a
changing colour temperature correction, as it requires a {\em smaller}
colour temperature correction than that in the standard high/soft state
(see \S 6). Thus either the disc inner
radius is {\em increasing}, or the corona is much more complex than
modelled here.

\begin{table*}
\centering
\begin{minipage}{175mm}
\caption{Best fit parameters obtained by a simultaneous ASCA/{\it RXTE} observation. 
Models are indicated with (a), (b), (c), (d), and (e). 
All the models consist of  MCD, {\sc thcomp} and power-law. 
In addition to these basic components, a narrow 
gaussian line and {\sl smedge} model are included in model (a). 
Models (b--e) include the full reflection cord in the {\it thcomp} model 
instead of adding the narrow gaussian and the {\sl smedge} model. 
In the case of model (c--e), the data in 5--12~keV is excluded from the spectral fit, $R_{\rm in}$ is fixed to 30~$R_{\rm g}$, 
and the value of $\xi$ is fixed to $10^2$, $10^3$, $10^4$, for models (c), (d), and (e), respectively. Normalization factors with the PCA data are used to calculate luminosities. 
}
\label{tab:asca-rxte}

\begin{tabular}{cccccccccc}
\hline \hline
model&$N_{\rm H}$ & $T_{\rm in}$ & $\Gamma_{\rm thc}$ & $T_{\rm e}$ & $L_{\rm disk}$ & $L_{\rm thc}$ & $L_{\rm pow}$&   smedge,line (a) & $\chi^2/{\rm dof}$\\
&$10^{21}{\rm cm^{-1}}$ & keV & & keV &($N_{\rm disk}$)  &($N_{\rm thc}$) &  & reflection (b--e) & \\
\hline \hline
(a)&$6.7\pm0.2$ & $0.54\pm0.02$&$2.34^{+0.02}_{-0.03}$ &  $14\pm2$  & 1.14 &2.87 &0.34 &   $E=7.5^{+0.3}_{-0.2}$ keV   &150.4/186\\
&& & && (35.0) &(57.9) &    &$\tau_{\rm max}=0.5^{+0.4}_{-0.1}$ &  \\
&& & && & & &   width$=4^{+5}_{-2}$~keV    &\\
&& & && & & &     $E=6.4\pm0.1$ keV &\\
&& & && & & &    EW=$43\pm12$ eV  &\\ 
 \hline
(b)&$7.1\pm0.2$ &$0.43^{+0.06}_{-0.01}$ &$2.74\pm0.03$ 
&$>140$ \footnote{The best fit value appeared in $>200$~keV.}
&$0^{<+0.26}$ 
&4.02/3.64 
& 0.15   & $\Omega/2\pi=1.7^{+0.2}_{-0.1}$    &215.2/188  \\
& & & &&($0^{<+10}$) &(112.4/111.0) &    &   $\xi=6\pm2$  &  \\
& & & && & &    &    $R_{\rm in}=12\pm2~R_{\rm g}$  & \\ \hline
(c)&$7.1^{+0.3}_{-0.2}$  &$0.53\pm0.03$ &$2.65^{+0.03}_{-0.02}$ &$36^{+26}_{-12}$ &$0.88 $  & 3.56/2.72 &0.25    & $\Omega/2\pi=1.7^{+>0.3}_{-0.4}$  &104.5/140   \\
& & & &&(27.6)  &(95.5/67.4) &    &   ($\xi=10^2$) &  \\ \hline
(d)&$6.5\pm0.2$&$0.59\pm0.02$ &$2.42\pm0.02$ &$17\pm4$ &$1.14 $  &2.85/2.30 &0.31    & $\Omega/2\pi=0.7\pm0.1$   &125.2/140   \\
& & & &&(32.3)  &(56.6/45.4) &    &   ($\xi=10^3$) & \\ \hline
(e)&$6.7\pm0.2$&$0.54^{+0.03}_{-0.02}$ &$2.37^{+0.01}_{-0.02}$ &$16\pm3$ &1.10  & 2.91/2.23 & 0.31  & $\Omega/2\pi=0.6\pm0.1$ &  120.6/140   \\
& & & &&(33.7)  &(57.2/45.3) &    &   ($\xi=10^4$) &  \\ 
\hline
 \end{tabular}
\end{minipage}
\end{table*}

\section{Simultaneous ASCA-PCA-HEXTE data}

One obvious caveat of the spectral modelling is that the low
temperatures inferred for the strong very high state mean that the
disc spectrum is not well covered by the PCA bandpass which starts
only at $\sim$3~keV. Hence we use the simultaneous ASCA GIS data to
extend the soft energy response. We use the same three continuum
component model and fitting conditions as above, except that the value
of $N_{\rm H}$ is now a free parameter.
Same as in \S4.2, except for a normalization factor, 
all the model parameters are constrained to be the same 
between the GIS, the PCA and the HEXTE data.
Table~\ref{tab:asca-rxte} shows the fitting results.
Figure~\ref{fig:asca} shows the raw data and the residuals between the
data and the model, and Fig.~\ref{fig:ascau} shows an unfolded
spectrum with the model. The model fits the simultaneous data very
well with $\chi^2/{\rm dof}=150.4/186$. The best fit value of $T_{\rm
in}= 0.54\pm0.02$~keV is much more tightly constrained, 
and its central value is slightly shifted to lower temperature than
$0.65^{+0.08}_{-0.04}$~keV derived from the
{\it RXTE} data alone.  

\begin{figure}
\centerline{\psfig{file=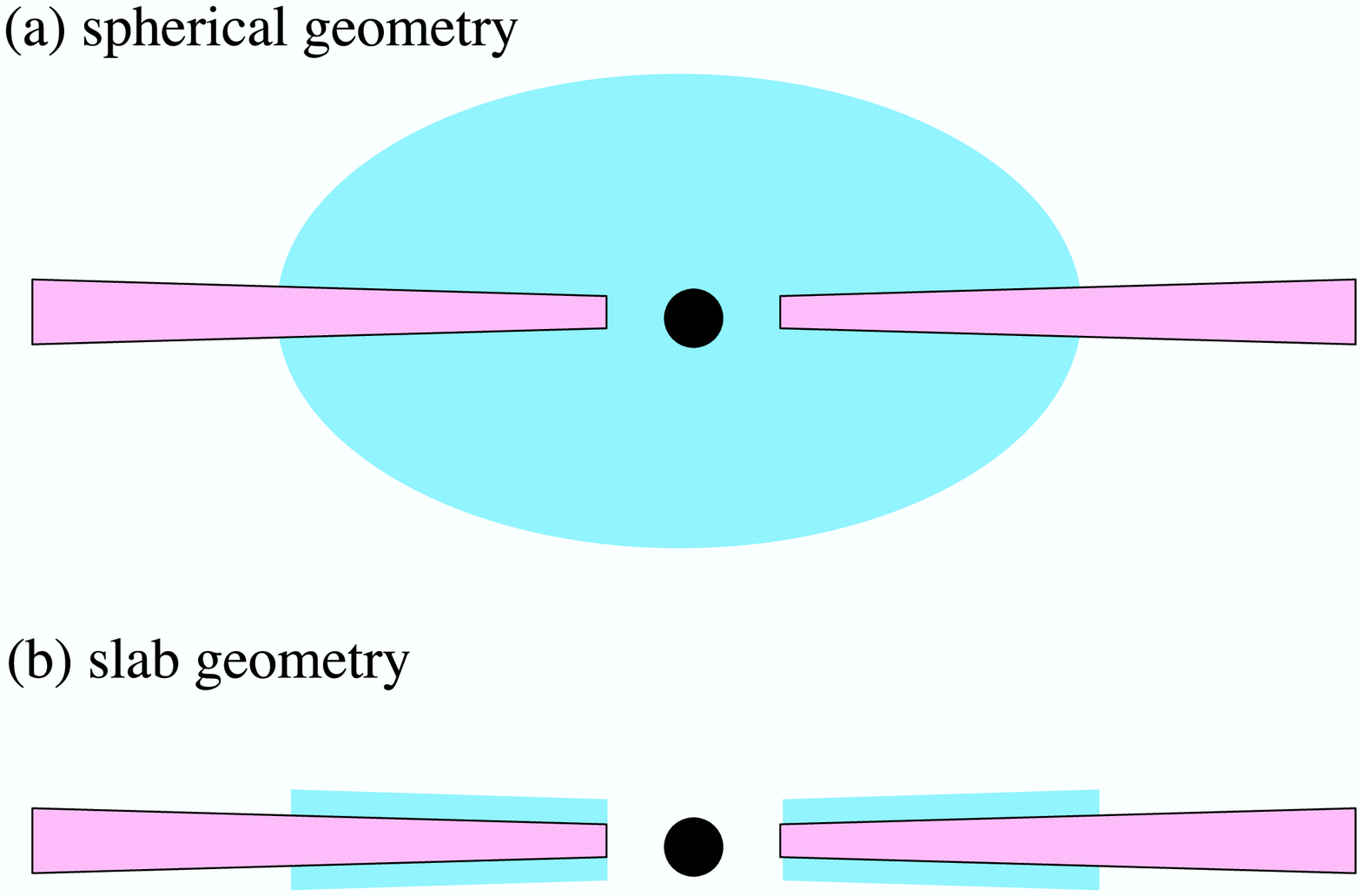,width=240pt}} 
\caption{Schematic pictures of assumed geometries of the corona. 
Spherical(a) and slab(b) geometries are shown.}
\label{fig:geo1}
\end{figure}

In order to check the results with a more physical model, we replace
the phenomenological line and smeared edge spectral features with the
full reflection code included in the {\sc thcomp} model (Zycki, Done
\& Smith 1998). We fix the
inclination and iron abundance at $70^\circ$ and solar abundance,
respectively.  These fit results are shown as model (b) in
Table~\ref{tab:asca-rxte}. The derived reflection parameters look
somewhat unphysical, with a large amount of mostly neutral
reflection. This is probably due to limitations of the reflection
modelling for an ionized disc in {\sc thcomp}, which is based on the
{\sc pexriv} reflected continuum calculations (Magzdiarz \&
Zdziarski 1995).  These codes calculate the ionisation balance in a very
simplistic way (Done \et  1992), assumming it to be constant
throughout the disc rather than varying as a function of height
(Nayakshin \et  2000; Ballantyne, Ross \& Fabian 2001). 

Another issue is that these models assume that Compton upscattering is
negligible below 12~keV.  However, for ionised material this can be
very important in determining the iron line and edge structure
observed (Ross, Fabian \& Young 1999).  While more accurate ionised
reflection models are available for spectral fitting (Nayakshin et
al. 2000; Ballantyne, Iwasawa \& Ross 2001), they assume a very simple continuum form
(power law, or power law with exponential cutoff) 
which means they cannot accurately describe the complex
spectral shape observed in these very high state data. Instead, we
estimate the impact of this uncertainty in the reflection modelling by
removing the iron line and edge region (5--12~keV) and refitting the
data with the {\sc thcomp} reflected continuum with an ionization
parameter, $\xi$, fixed at $10^2$, $10^3$ and $10^4$, (models (c),
(d) and (e) respectively in Table 2).

The disc luminosities and temperatures derived from the series of
models (a--e in Table 2) are plotted as filled black squares on the
$T-L$ plots (Fig.~\ref{fig:t-l}b--d). These show the range of
systematic uncertainties present from the spectral modelling. However,
{\em all} of the fits detailed in Table 2 have an even lower disc
temperature than that derived from fits to the PCA+HEXTE data, but
with a similar luminosity. These show even more clearly that the
strongly Comptonised spectra are {\em not} easily consistent with an
untruncated disc. For the simple corona 
geometries of Fig.~\ref{fig:geo1}, then in order to explain 
the high inferred intrinsic disc luminosity,
$L_{\rm disk}^{\rm int}\sim2$--$3\times10^{38}~{\rm erg~s^{-1}}$, with the
same disc radius as seen in the standard high/soft state requires 
$T_{\rm in} >$~0.85--0.9~keV. By contrast, the 99.9\% upper limits 
($\chi^2/{\rm dof}=1.35$ for dof of 186--188)
of the disc temperature from the ASCA-PCA-HEXTE data is $<0.66$~keV for all
the models (a--e) in Table 2. 
Fixing $T_{\rm in} =0.85$~keV results in 
completely unacceptable fits ($\chi^2/\rm dof$ =4.6--5.6).

More conservatively, even 
when the systematic errors of the PCA data and the HEXTE data 
increased to 1~\%, the 99.9~\% upper limit is still lower than 0.69~keV.
Since deviations between the modeled and measured Crab spectrum 
do not exceed 1~\% as noticed by many authors including Revnivtsev \et ~(2003), $T_{\rm in}=0.69$~keV can be considered as the 
securest upper limit of the disc inner temperature.
%
It is clear that the low disc temperature derived
in the strong very high state spectra from the PCA+HEXTE fits are {\em not}
due to lack of PCA coverage of the disc spectrum, or the
spectral modelling of the reflection emission. 

\begin{figure}
\centerline{\psfig{file=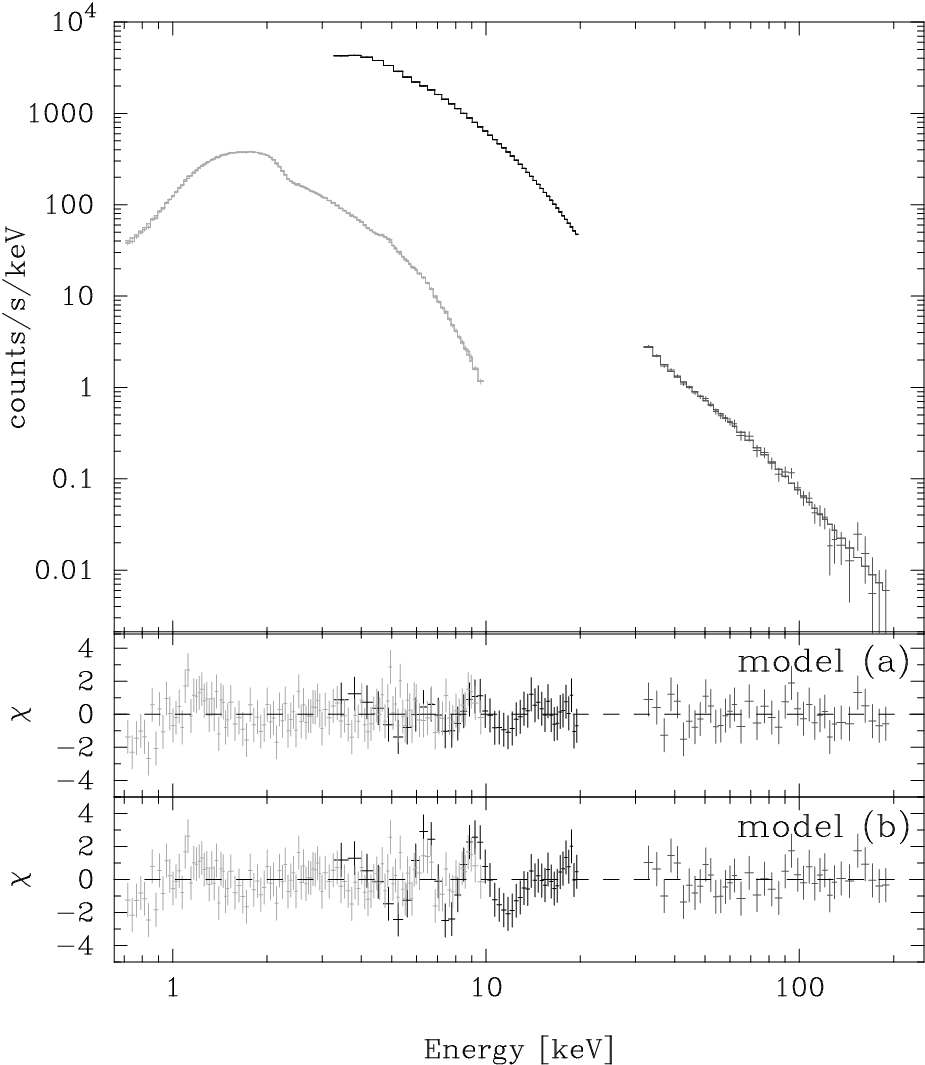,width=240pt}}
\caption{The simultaneous spectral fit on GIS, PCA and HEXTE, with the best fit 
model predictions.}
\label{fig:asca}
\centerline{\psfig{file=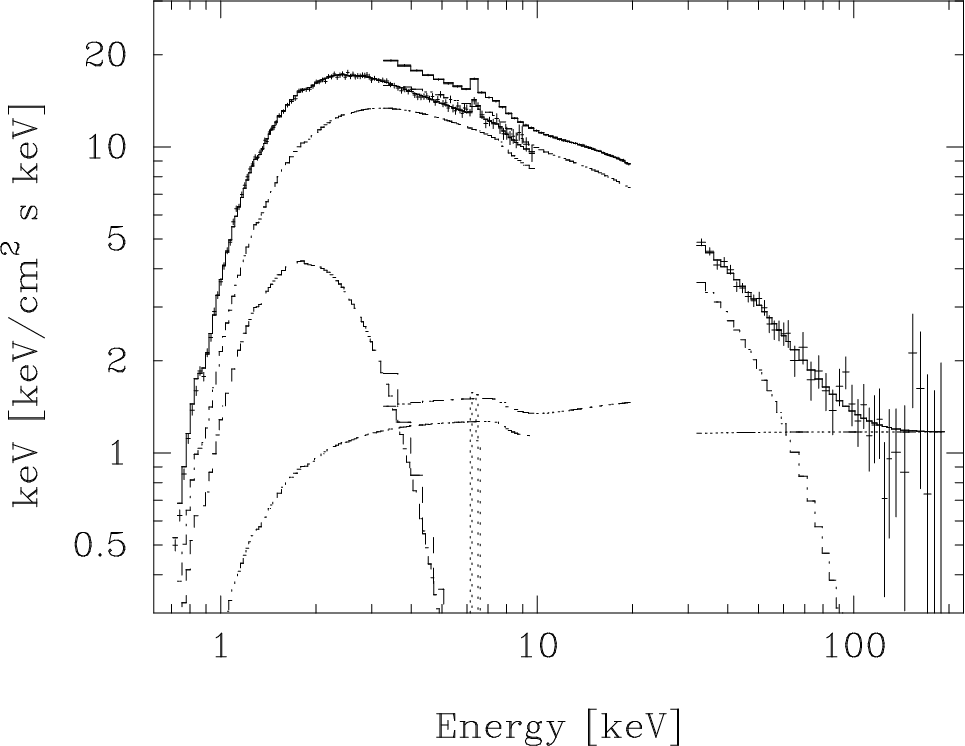,width=240pt}}
\caption{Unfolded spectrum of Fig.~\ref{fig:asca} with the predictions of model (a).}
\label{fig:ascau}
\end{figure}

\section{Discussion}

We have shown that the strong very high state spectra really do
have a rather low temperature disc for its luminosity. These observations
suggest that the disc is truncated at larger 
radii than the last stable orbit in the
strong very high state. However, this is based on the assumptions that
(1) the correction factor of color to effective temperature is kept
approximately constant, and  
(2) the corona fully covers the disc as illustrated in
Fig.~\ref{fig:geo1}

\subsection{Change of color temperature correction}

Figure~\ref{fig:evolution2}c represents the time history of the
apparent inner radius $r_{\rm in}$, as derived from the temperature
and luminosity values given in Fig.~\ref{fig:t-l}d. The true inner
radius $R_{\rm in}$ is related to $r_{\rm in}$ as $R_{\rm in}\propto
\kappa^2 r_{\rm in}$, where $\kappa$ is the colour temperature
correction.  For constant $\kappa$, the observed variability of
$r_{\rm in}$ is just caused by change of the inner radius.  However, a
change in the colour temperature correction could be expected as a
function of disc luminosity (Merloni \et  2000; GD04; Kawaguichi 2003), and
illumination of the upper layers of the disc (Nayakshin \et  2000)
and/or conductive heating could also lead to an increase in $\kappa$.

Close inspection of Fig.~\ref{fig:t-l}d shows that the 
data points corresponding to the weak very high state (anomalous regime)
are slightly lower than the solid line denoting constant $r_{\rm in}$. 
Although this is not very significant, it could perhaps indicate a
constant radius disc with slightly higher colour temperature correction. 
Conversely, the disc in the strong very high state requires a {\em decrease}
in colour temperature correction by a factor of $\sim 60$\%
i.e. $\kappa\sim 1$. This would imply that the observed colour
temperature was the same as the effective disc temperature, which is
only possible when true absorption opacity dominates over electron
scattering. This {\em cannot} be the case at the observed
$\sim$0.5~keV temperatures. Changing $\kappa$ cannot 
explain the strong very high state spectra with the disc with a
constant radius. 

The above discussion assumed that the disc spectrum can be described
by a simple colour temperature correction. Plainly this may not be the
case, and the disc spectrum could be completly distorted, for example
if the inner disc colour temperature correction were much higher than 
at outer radii. The spectrum would then look like a low temperature disc
with some fraction of the emission at much higher temperature. This is 
exactly what these strong very high state spectra look like, so we
examine the effect of Comptonization as a function of radius on the disc
spectrum below. 

\subsection{Geometry and energetics of the corona}

So far, we assumed the corona fully covers the disc as illustrated in
Fig.~\ref{fig:geo1}.  However several other geometries can be
considered, as illustrated in Fig.~\ref{fig:geo3}.  
If the corona is out of our line of sight to the disc e.g. a small
inner region, perhaps physically representing an inner jet or bulk
Comptonisation of the infalling material (see Fig.~\ref{fig:geo3}a and
b), then the Compton scattering does not affect the observed disc
flux, so the correction used above {\em overestimates} the intrinsic
disc emission.  The problem of such a geometry
is that the Comptonising region
would see many fewer disc photons than an observer, in conflict with
the data (see Tables 2 and 3). Even without this problem, the best
that such a geometry could do is reduce the estimated intrinsic disc
luminosity to that observed (i.e. where none of the disc photons in
our line of sight are intercepted). This puts us straight back to
Fig.~\ref{fig:t-l}b, where the ASCA+PCA+HEXTE data show a much lower
temperature disc than expected for its luminosity in the very high
state.

Alternatively, the corona could
be patchy (Fig.~\ref{fig:geo3}c), or cover only the inner disc
(Fig.~\ref{fig:geo3}d). In both these geometries the corona
is in the line of sight to the disc, so again this predicts that the
Compton scattered photons we see are removed from the observed disc
emission. The intrinsic disc emission should still be estimated as in
Fig.~\ref{fig:t-l}d. While the patchy corona could take
photons uniformly from the whole disc, the inner corona
(Fig.~\ref{fig:geo3}d) would preferentially take only the inner disc
(highest temperature) photons, perhaps leading to an underestimate of
the disc temperature. This is analogous to having a variable 
colour temperature correction with radius (see above). 
We model this by switching the seed photons in
{\sc thcomp} to blackbody (rather than the MCD), and allow the
temperatures of the observed inner disc and seed photons to be
different. This gives a very strong upper limit on the seed photon
temperature for the ASCA-PCA-HEXTE data of $<0.6$~keV, showing that
the low disc temperature is {\em not} an artifact of the seed photon
distribution assumed. 

However, the existence of the corona {\em changes} the disc temperature
structure. In the limit of a corona that covers the whole disc then
this has little effect as both the disc luminosity and temperature are
reduced (Svensson \& Zdziarski 1994), but for an inner disc corona
the behaviour is more complex. Only the inner disc temperature and
luminosity are reduced, but these form the seed photons for the
Compton scattering. For a fraction $f$ of the total accretion power
dissipated in the corona, the inner disc temperature is reduced by a
factor $(1-f)^{1/4}$ (Svensson \& Zdziarski 1994). This is an 
underestimate of the inner disc temperature as it intercepts and 
reprocesses a large fraction (up to a half) of the coronal flux.
We will do more detailed modelling of this in a subsequent paper, 
but estimate that the size of this effect is not enough
to make the temperature and luminosity consistent with the standard high/soft
state radius disc.

\section{A scenario for the disc evolution in the 1998 outburst of XTE~J1550-564}
\subsection{A scenario derived by spectral studies}

The discussion above shows that there are no easy alternatives to the
conclusion that the inner disc is truncated in the strongly
Comptonised very high state spectra. If so, then there must be some
overlap between the inner hot region and the truncated disc in order
for it to intercept enough seed photons, but the larger radius disc
can trivially produce the lower observed disc temperature at high
luminosity.  This geometry is rather similar to the truncated disc/hot
inner flow inferred for the low/hard state, so would give a physical
basis for the well known similarities between the very high state
spectra and the intermediate state spectra (seen towards the very end
of the low/hard state; Belloni \et  1996; Mendez \et  1997). 
It could also explain the lower
frequency QPO seen in the very high state (e.g. the review by van der
Klis 2000; di Matteo \& Psaltis 1999).

A schematic picture for the whole disc evolution during the 1998
outburst of XTE~J$1550-564$ is shown in Fig.~\ref{fig:geo2}.  The
outburst starts in the low/hard state, with a truncated disc, and hot
inner flow (panel a). 
The increase in mass accretion rate increases
the optical depth of the inner flow before the inner disc has time to
form, leading to an optically thick, cooler inner flow and truncated
disc (the strongly Comptonised very high state, panel b) on $\sim$day
5 (WD01).  
During days~5--30, the fraction of energy dissipated in the corona to that from the 
disc is kept almost constant (Fig.~\ref{fig:evolution2}f), while $T_{\rm e}$ and 
$\Gamma_{\rm thc}$ of the corona changed 
with time (Fig.~\ref{fig:evolution2}d and e). 
Through this period, the X-ray flux reached at the peak (day~13). 
The radio flare was observed at this time with a continuance of 5~days, 
and VLBA image showed evolving structure (Hannikainen \et ~2001).

During days 30--40 the inner disc is able to form at
progressively smaller radii, perhaps due to the correlated decrease in
fractional power dissipated in the corona (see
Fig.~\ref{fig:evolution2}f) which reduces the coronal
heating/irradiation. The disc finally reaches the last stable orbit at
around day~40, causing a noticeable change in disc temperature behaviour 
despite little change in luminosity (Fig.~\ref{fig:evolution2}b). 
The power dissipated in the corona continues to drop,
leading to more weakly Comptonised spectra above the untruncated disc
(the weak very high state, anomalous regime; days 40-52). 
This eventually becomes a small ($<$~20\%) fraction of
the total power, giving the standard high/soft state (after day 52, panel e).
This is similar to the geometry for the spectral states 
proposed by Esin \et (1997), except that
the strongly Comptonised very high state spectra corresponds here to 
a truncated disc. 

While this gives a good overall description of the properties of
XTE~J$1550-564$ during its outburst, the remaining problem is what
makes the truncated disc with the strong Compton corona.  Advection
dominated accretion flows cannot be maintained at such high luminosities
(e.g. Esin \et 1997), although these are only an approximation to the
complex nature of the optically thin, hot flow which is predicted
from numerical simulations (Hawley \& Balbus 2002). Physically,
such truncation could arise from disc overheating (Beloborodov 1998),
or by strong conduction/irradiative heating from the corona, or,
perhaps most attractively, by 
disruption of the inner disc by jet formation.  However,
the total mass accretion rate cannot be the trigger for this behaviour
since (apart from the peak) the very high state spectra have total
luminosity which is similar to that in the disc dominated 
regime seen in days 100--180.  Therefore
some other parameters are required to drive the structural changes in
the disc (see e.g. van der Klis 2001; Homan \et ~2001; MR03)


\begin{figure}
\centerline{\psfig{file=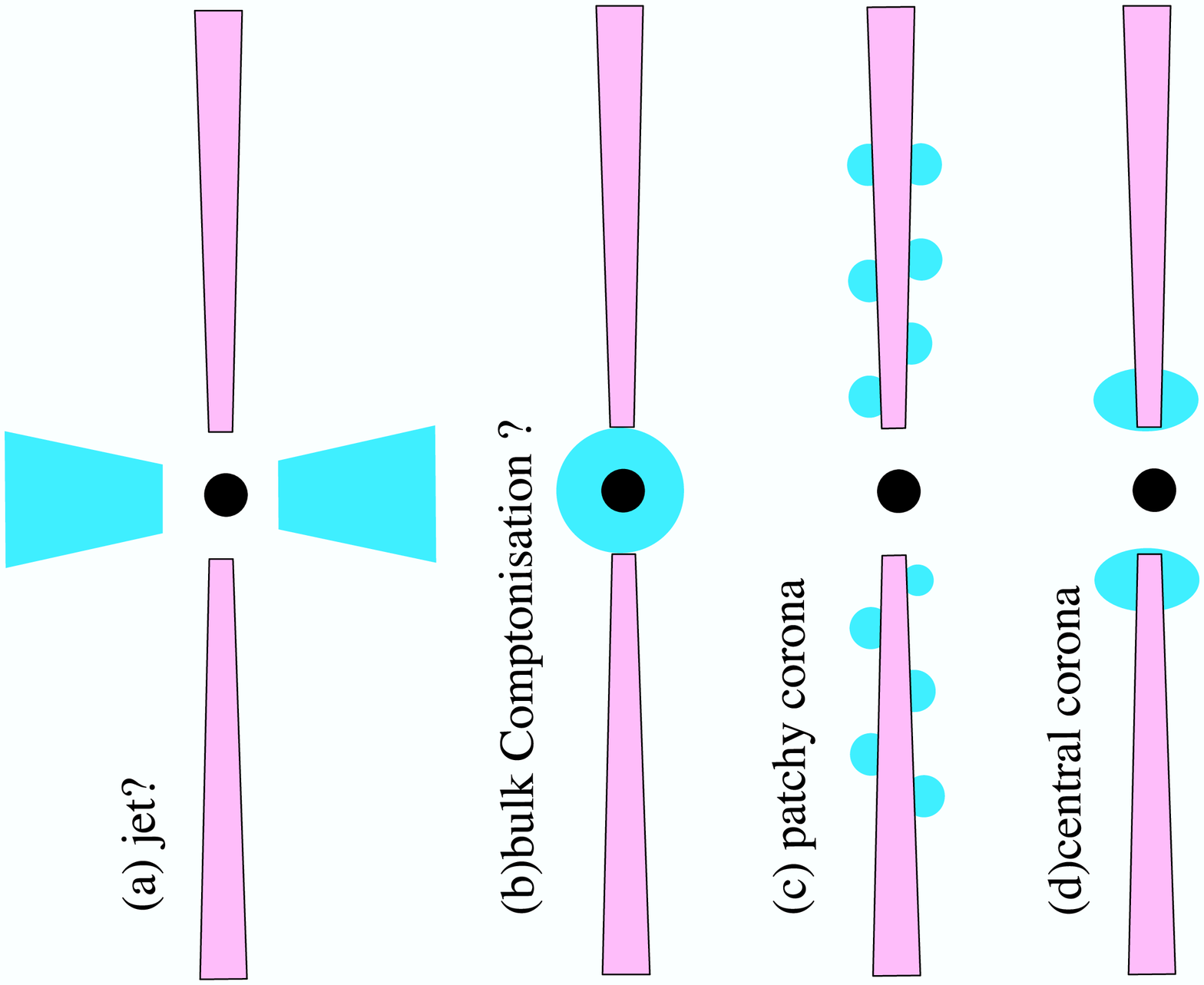,width=240pt,angle=-90}} 
\caption{Other candidates of geometries of the material which up-scatters the disc photons.}
\label{fig:geo3}
\end{figure}

\begin{figure}
\centerline{\psfig{file=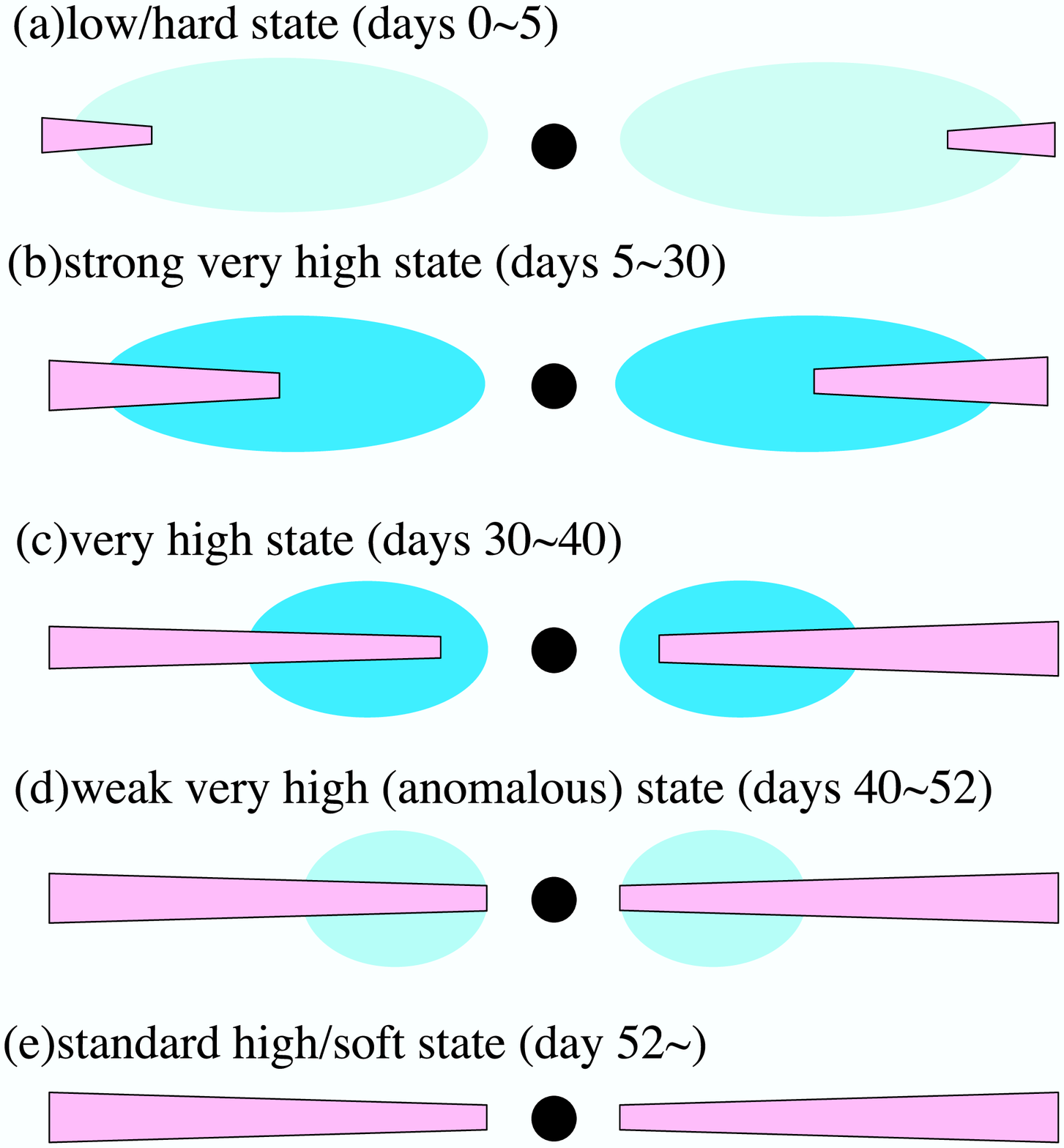,width=240pt,angle=-90}} 
\caption{Schematics of obtained picture of the disc evolution on the beginning of the first 
outburst of XTE~J$1550-564$. 
Panels (a), (d) and (e) refer to pictures of 
"Low State", "High State" and  "Very High State" in Figure~1 of Esin \et  (1997),
respectively.}
\label{fig:geo2}
\end{figure}

\subsection{Consistency with the QPO behavior}

The type of QPO seen is roughly consistent with the 
changes in geometry shown in Fig. 10, and with the
spectral state (Table~1)
as suggested by Homan \et ~(2001) based on the spectral hardness ratio. 
Since the QPO frequencies are thought to reflect 
the size of the Comptonising region and the disc inner radii, 
it is meaningful to compare evolution of the QPO frequencies 
with that of disc inner radii estimated by the spectral analyses. 
As we can see in Fig.~\ref{fig:asm}, through day~18--30, the
frequency of the low-frequency QPO increased from $\sim2.7$~Hz, 
and after the day~30, it was almost kept constant at $\sim6$~Hz. 
This is very consistent with the decrease in  $r_{\rm in}$ derived from the 
spectra through days 5--40 as 
shown in Fig.~\ref{fig:evolution2}c.
Indeed, Homan \et ~2001 suggested almost the same scenario of evolution of the 
accretion disc based on the QPO behavior.

However, there is some inconsistency between the timing and the spectral 
behavior before day~18.  On one hand, the low-frequency QPO increased to 
$\sim10$~Hz around the peak (day~13). This frequency is {\em higher} than 
the $\sim6$~Hz which was observed after days~30--40, suggesting that the disc
extends closer to the black hole during the peak. Yet the X-ray spectra 
imply that the disc extends down to the last stable orbit during days 30--40
and truncates at {\em larger} radii during the peak(Fig.~\ref{fig:evolution2}c). 
Moreover,  the high-frequency QPO which is sometimes seen in the very high state
data (Fig.~\ref{fig:asm}) shows rather different behaviour. 
In Fig.~\ref{fig:t-l}, we marked the data points which have high-frequency QPOs 
with big open diamonds. Some green open squares showed both the characteristics of 
large inner radius by the spectral fits and the high frequency of the QPOs. 
Therefore, the QPO and the spectral behavior are not easily consistent, pointing
either to a lack of understanding of how QPOs and/or disc spectra relate to the 
inner radius.

As a brief comment, the strong very high state data with high frequency QPO and 
the low-frequency QPOs of the higher frequencies may be considered to be relate to 
the jet ejection.
As described in \S~4.2, the radio flux was observed to increase suddenly at day 13, 
the peak of X-ray flux (Wu \et ~2002), and the 
extending radio jet was reported on day 15 by Hannikainen \et ~(2001).
In Fig.\ref{fig:t-l}, the data points 
which show strong radio emission were also marked with 
big circle plus cross. Thus,
the data with high-frequency QPO can be understood to have some relation to 
the jet ejection. 

\section{Summary and Conclusions}

The disc dominated high/soft state of Galactic black holes can be well
modelled by a disc of constant inner radius and colour temperature
correction. This most probably indicates that the disc extends down to
the innermost stable orbit around the black hole. This is clearly seen
in the 1998 outburst of XTE~J$1550-564$. However, the strongly comptonised
very high state spectra seen at the start of the outburst do not follow this trend.
Careful modelling of these 
spectra show that the disc temperature is somewhat lower than expected
for the observed disc luminosity for a constant radius
disc. Correcting the disc luminosity for the effects of Compton
scattering in a disc-corona geometry make this discrepancy much more
marked. We discuss the effects of changing colour temperature
corrections, and different geometries, but none of these can
convincingly reproduce the observed disc temperatures and
luminosities. We will model the effects of a more complex coronal
geometry (e.g. one with strong radial gradients in optical depth:
Fig.~\ref{fig:geo3}d) in a subsequent paper, but the simplest solution
is that the inner disc is truncated in the strong very high state.

\subsection*{Acknowledgements}

The authors would like to thank an anonymous referee for his/her detailed comments.
The present work is supported in part by a 
Sydney Holgate fellowship in Grey College, University of Durham, 
JSPS Postdoctoral Fellowship for Young Scientists, 
and by JSPS grant of No.13304014. 
A.K. is supported by special postdoctoral researchers program in RIKEN.

\end{document}